\documentclass[a4paper,11pt]{article}
\usepackage{a4wide,amsmath,amssymb,bbm}
\usepackage[english]{babel}

\usepackage{color}

\usepackage{hyperref,enumerate}
\hypersetup{
    colorlinks,
    linkcolor={black},
    citecolor={black},
    urlcolor={black}
}

\renewcommand{\u}{\underline}
\newcommand{\tr}{\mathrm{tr}}
\newcommand{\Ad}{\mathrm{Ad}}
\newcommand{\be}{\begin{equation}}
\newcommand{\ee}{\end{equation}}
\newcommand{\alg}{\mathfrak}
\makeatletter

\@addtoreset{equation}{section} \makeatother

\begin{document}

\hfill{NORDITA 2018-041}

\vspace{30pt}

\begin{center}
{\huge{\bf Non-abelian T-duality and Yang-Baxter deformations of Green-Schwarz strings}}

\vspace{80pt}

Riccardo Borsato$\, ^a$ \ \ and \ \ Linus Wulff$\, ^b$

\vspace{15pt}

{
\small {$^a$\it Nordita, Stockholm University and KTH Royal Institute of Technology,\\ Roslagstullsbacken 23, SE-106 91 Stockholm, Sweden}\\
\vspace{5pt}
\small {$^b$\it Department of Theoretical Physics and Astrophysics, Masaryk University, 611 37 Brno, Czech Republic}
\\
\vspace{12pt}
\texttt{riccardo.borsato@su.se, wulff@physics.muni.cz}}\\

\vspace{100pt}

{\bf Abstract}
\end{center}
\noindent
We perform non-abelian T-duality for a generic Green-Schwarz string with respect to an isometry (super)group $G$, and we derive the transformation rules for the supergravity background fields. Specializing to $G$ bosonic, or $G$ fermionic but abelian, our results reproduce those available in the literature. We discuss also continuous deformations of the T-dual models, obtained by adding a closed $B$-field before the dualization. This idea can also be used to generate deformations of the original (un-dualized) model, when the 2-cocycle identified from the closed $B$ is invertible. The latter construction is the natural generalization of the so-called Yang-Baxter deformations, based on solutions of the classical Yang-Baxter equation on the Lie algebra of $G$ and originally constructed for group manifolds and (super)coset sigma models. We find that the deformed metric and $B$-field are obtained through a generalization of  the map between open and closed strings that was used also in the discussion by  Seiberg and Witten of non-commutative field theories. When applied to integrable sigma models these deformations preserve the integrability.

\pagebreak 
\tableofcontents

\setcounter{page}{1}


\section{Introduction}
While ordinary abelian T-duality is an exact symmetry of string perturbation theory, its non-abelian generalization~\cite{delaOssa:1992vci} is not \cite{Giveon:1993ai,Alvarez:1993qi}. It should be rather viewed as a solution-generating technique in supergravity, since it (typically) maps one string background to another, inequivalent one. Starting with the work of \cite{Sfetsos:2010uq}, which gave a prescription for the transformation of the RR fields, it has been successfully applied to construct several interesting supergravity solutions, e.g. \cite{Lozano:2012au,Itsios:2012zv,Itsios:2013wd,Lozano:2016kum,Lozano:2016wrs,Lozano:2017ole}.

Like its abelian version, non-abelian T-duality (NATD) can be understood as a canonical transformation~\cite{Lozano:1995jx,Sfetsos:1996pm,Sfetsos:1997pi}, so that the dualization preserves the (classical) integrability of the sigma model (when present). To be more precise, starting from a sigma model whose equations of motion are equivalent to the flatness of a Lax connection, one obtains a dual model whose equations of motion can also be put into Lax form.  Here we want to exploit this property in order to generate integrable deformations of sigma models, following the ideas of~\cite{Hoare:2016wsk,Borsato:2016pas,Borsato:2017qsx}.\footnote{Another class of integrable deformations related to NATD are the so-called $\lambda$-deformations of \cite{Sfetsos:2013wia,Hollowood:2014rla,Hollowood:2014qma}.} The deformations are interesting also because they (partially) break the initial isometries. We remark that integrability is not essential for the construction, and the deformations can be carried out also for non-integrable models. Some of the deformations constructed here may be viewed as continuous interpolations between the ``original'' model and the ``dual'' one obtained after applying NATD.

Starting from a \emph{generic} type II Green-Schwarz superstring whose isometries contain a (super)group $G$, we work out the transformation rules for the supergravity background fields under NATD with respect to $G$. The derivation is performed in section~\ref{sec:NATD}, where all orders in fermions are taken into account by working in superspace. When choosing a bosonic $G$ and focusing on the bosonic supergravity fields, the transformation rules reproduce those of~\cite{Sfetsos:2010uq,Lozano:2011kb}, including the Ramond-Ramond (RR) fields whose transformations were conjectured by analogy with the abelian case~\cite{Hassan:1999bv}. Moreover, when the Lie algebra of $G$ consists of only (anti)commuting fermionic generators, we also reproduce the rules for fermionic T-duality derived in~\cite{Berkovits:2008ic} from the pure spinor string.
As expected, we show that after NATD one still obtains a kappa symmetric Green-Schwarz superstring. It follows from~\cite{Wulff:2016tju} that the target space is therefore a solution of the generalized supergravity equations of~\cite{Arutyunov:2015mqj,Wulff:2016tju}. 
When $G$ is unimodular (i.e. the structure constants of its Lie algebra satisfy $f^J_{IJ}=0$) the background fields satisfy the standard type II supergravity equations, and the (dualized) sigma model is Weyl invariant. When $G$ is not unimodular there is typically an anomaly which breaks Weyl invariance and obstructs the interpretation of the dual model as a string~\cite{Alvarez:1994np,Elitzur:1994ri}. We will also discuss exceptions to this, given by the ``trivial solutions'' of~\cite{Wulff:2018aku}.

Deformations of the non-abelian T-dual backgrounds may be generated by adding a closed $B$-field before dualizing. The deformation will be controlled by one or more continuous parameters that enter the definition of this $B$. From the point of view of the original model, adding a $B$-field with $dB=0$ does not affect the local physics, since this term does not change the equations of motion. We will nevertheless obtain a non-trivial deformation and a dependence on $B$ in the equations of motion after applying NATD, since this transformation involves a \emph{non-local} field redefinition.\footnote{If $B$ is not just closed but also exact, it contributes to the action of the original model as a total derivative and it can be dropped. Even if kept, the dependence on this $B$ can be removed by a (local) field redefinition even after applying NATD. Therefore an exact $B$ generates a trivial deformation of the dual model.} 
Writing $B=\frac12(g^{-1}dg)^J\wedge(g^{-1}dg)^I\, \omega_{IJ}$ with $g\in G$, the condition $dB=0$ is equivalent to $\omega$ being a 2-cocycle on the Lie algebra of $G$. The resulting models were dubbed deformed T-dual (DTD) models in~\cite{Borsato:2016pas}, and we refer to section~\ref{sec:DTD} for more details.

In~\cite{Borsato:2016pas,Borsato:2017qsx} it was proved that a DTD model constructed from a principal chiral model (PCM) or supercoset sigma model with $\omega$ invertible is actually equivalent (thanks to a local field redefinition) to the so-called Yang-Baxter (YB) sigma models~\cite{Klimcik:2002zj,Klimcik:2008eq,Delduc:2013fga,Kawaguchi:2014qwa,Matsumoto:2015jja,vanTongeren:2015soa} based on an $R$-matrix solving the classical Yang-Baxter equation.\footnote{These are sometimes called ``homogeneous'' YB models. In the ``inhomogeneous'' YB models $R$ solves the \emph{modified} classical Yang-Baxter equation. They were first introduced in~\cite{Klimcik:2002zj,Klimcik:2008eq} and later generalized to the supercoset case in~\cite{Delduc:2013qra}, where the so-called $\eta$-deformation of $AdS_5\times S^5$ was constructed. The inhomogeneous YB models are not related in such a simple way to NATD and we will not consider them further here.} The $R$-matrix is related to the 2-cocycle simply as $R=\omega^{-1}$. The equivalence was first proposed and checked on various examples in~\cite{Hoare:2016wsk}.\footnote{An equivalent construction, applying NATD on a centrally extended algebra, was used there. See also~\cite{Hoare:2016wca}.}
When the $R$-matrix acts only on an abelian subalgebra YB deformations are simply TsT (T-duality - shift - T-duality) transformations~\cite{Osten:2016dvf}, so that we can think of YB deformations as the ``non-abelian'' generalization of TsT transformations. 
Here we propose to use the connection to NATD in order to extend the applicability of YB deformations, from just PCM and supercoset models to a generic sigma model with isometries. We do this in section~\ref{sec:YB} by carrying out the field redefinition which leads from the DTD model to the YB model in the case of invertible $\omega$. Although the construction comes from a deformation of the dual model, when sending the continuous deformation parameter of the YB model to zero we recover the \emph{original} model. These deformations may be particularly interesting for the AdS/CFT correspondence, and in section~\ref{sec:brane} we use our results to ``uplift'' a YB deformation of $AdS_5\times S^5$ -- that cannot be interpreted as (a sequence of) TsT transformations -- to a deformation of the full D3-brane background, of which $AdS_5\times S^5$ is the near-horizon limit. 

For YB deformations of the PCM or (super)coset models, it is easy to see that the background metric and $B$-field are related to the metric of the original model by  a map that coincides with the open/closed string map used also by Seiberg and Witten in~\cite{Seiberg:1999vs}. For YB deformations the open string non-commutativity parameter is identified with the $R$-matrix itself~\cite{Araujo:2017jkb}. Based on this observation it was suggested in~\cite{Bakhmatov:2017joy} that  this map could be used to generate solutions to (generalized) supergravity.\footnote{In~\cite{Bakhmatov:2018apn} it was shown that the map generates solutions of the generalized supergravity equations if the non-commutativity parameter satisfies the classical Yang-Baxter equation.} Our results, based on the construction of \cite{Borsato:2016pas}, generalize this to cases with a non-vanishing $B$-field in the original model. Our derivation also ensures that the YB backgrounds are automatically solutions of the (generalized) supergravity equations.
Yet another approach to such general (homogeneous) YB deformations was proposed in the context of doubled field theory, since known YB deformations were shown to be equivalent to so-called $\beta$-shifts~\cite{Sakamoto:2017cpu,Lust:2018jsx,Sakamoto:2018krs}. In section~\ref{sec:AdS3-ex} we check in an example that a recent solution generated in~\cite{Sakamoto:2018krs} coincides with the one obtained from our method based on NATD.

In the next section we collect the transformation rules for the background fields under NATD and under a generic YB deformation.

\section{Summary of the transformation rules}\label{sec:summary}
In this section we wish to present and summarize in a self-contained way the transformation rules derived in the paper, so that the reader may consult them without the need of going through the whole derivation. 

\subsection{Rules of (bosonic) NATD}\label{sec:summary-NATD}
Here we summarize the NATD transformation rules for the bosonic supergravity fields only, when we take $G$ to be an ordinary (i.e. non-super) Lie group. The general transformations can be found in section~\ref{sec:NATD} (for the case of $G$ a supergroup see footnote \ref{foot:super}).
It is convenient to rewrite the background fields in a way that makes the $G$ isometry manifest. The metric, for example, will be written in the following block form
\be
G_{\mu\nu} =\left(\begin{array}{cc}
G_{mn}&G_{mj}\\
G_{i n}&G_{ij}
\end{array}
\right)\,,\qquad
G_{i n}=\ell_i^I G_{In},\qquad
G_{ij}=\ell_i^I\ell_j^JG_{IJ}\,.
\ee
We have chosen coordinates such that we can split indices into $(i,m)$, where $i$ takes $\dim G$ values and $m$ labels the remaining spectator fields which do not transform under $G$. We have collected our conventions in appendix~\ref{sec:conv}. It is also convenient to rewrite certain blocks by extracting $\ell_i^I$, defined by $g^{-1}\partial_i g=\ell_i^IT_I$, where $g\in G$ and $I=1,\ldots,\dim G$ is an index in $\alg g$ (the Lie algebra of $G$) so that $[T_I,T_J]=f_{IJ}^KT_K$. The dependence on the coordinates $x^i$ (i.e. the coordinates to be dualized) is all  in $\ell_i^I$, so that $G_{IJ},G_{Im},G_{mn}$ only depend on the spectators $x^m$. The transformation rules will be presented in terms of these objects, and we will continue to call them ``metric'' and ``$B$-field'' also when writing them with indices $(m,I)$ instead of $(m,i)$. In order to have a uniform derivation and presentation, we do not restrict further the range of the index $I$ even when a local symmetry is present.\footnote{Therefore the range of $(m,I)$ can exceed ten. Both the original and the final action are still written only in terms of ten physical coordinates thanks to the local symmetry that survives NATD and removes the additional degrees of freedom, see the discussion in section~\ref{sec:local}.} We refer to section~\ref{sec:NATD} for more details.
 Setting fermions to zero the transformation rules for the metric and $B$-field in (\ref{eq:Etildepm}--\ref{eq:Eatilde}) read\footnote{The coordinates $\nu_I$ that result from the dualization naturally have lower indices, since they parameterize the dual space. To have the standard upper placement of indices also in the dualized model we declare that those indices are raised with the Kronecker delta $\nu^I=\delta^{IJ}\nu_J$, and the total set of coordinates is ($x^m,\nu^I$). }
\begin{align}
&\tilde G_{mn}=G_{mn}-\big[(G-B)N(G-B)\big]_{(mn)}\,,\label{eq:tildeGNATD}\\
&\tilde G_{mI}=\tfrac12\big[(G-B)N\big]_{mI}-\tfrac12\big[N(G-B)\big]_{Im}\,,\qquad
\tilde G_{IJ}=N_{(IJ)}\,,\nonumber
\\
\nonumber\\
&\tilde B_{mn}=B_{mn}+\big[(G-B)N(G-B)\big]_{[mn]}\,,\\
&\tilde B_{mI}=-\tfrac12\big[(G-B)N\big]_{mI}-\tfrac12\big[N(G-B)\big]_{Im}\,,\qquad
\tilde B_{IJ}=-N_{[IJ]}\,,\nonumber
\end{align}
where $N_{IJ}=\delta_{IK}N^{KL}\delta_{LJ}$ etc. and
\begin{equation}
N^{IJ}=\left(G_{IJ}-B_{IJ}-\nu_Kf^K_{IJ}\right)^{-1}\,.
\label{eq:NIJ}
\end{equation}
The transformation of the RR fields, encoded in the bispinor (for more on the conventions see \cite{Wulff:2013kga,Wulff:2016tju})
\begin{equation}
\mathcal S^{12}=\left\{
\begin{array}{cc}
\mathcal F^{(0)}-\tfrac12\mathcal F^{(2)}_{ab}\Gamma^{ab}+\tfrac{1}{4!}\mathcal F^{(4)}_{abcd}\Gamma^{abcd} & \mbox{IIA}\\
-\mathcal F^{(1)}_a\Gamma^a-\tfrac{1}{3!}\mathcal F^{(3)}_{abc}\Gamma^{abc}-\tfrac{1}{2\cdot5!}\mathcal F^{(5)}_{abcde}\Gamma^{abcde} & \mbox{IIB}
\end{array}
\right.\,,
\end{equation}
given in (\ref{eq:S12tilde}) is given by the action of a Lorentz transformation $\Lambda\in O(1,9)$ as
\begin{equation}
\tilde{\mathcal S}^{12}=\hat\Lambda\mathcal S^{12}\,,\qquad\Lambda^{ab}=\eta^{ab}-2E_I{}^aN^{IJ}E_J{}^b\,,
\end{equation}
where $G_{IJ}=E_I{}^aE_J{}^b\eta_{ab}$, and we denote by $\hat\Lambda$ the Lorentz transformation acting on spinor indices that multiplies $\mathcal S^{12}$, defined such that $\Lambda^a{}_b\Gamma^b= \hat\Lambda^T\Gamma^a\hat\Lambda$.
Finally the generalized supergravity fields $K$ and $X$ given in (\ref{eq:Ktilde}--\ref{eq:Xtilde}) become\footnote{Here we drop the tilde since these fields are not present before dualization. Also note that we have raised the index on $K$ with $\tilde G^{-1}$ in order to get a simpler expression. We assume the original dilaton $\phi$ to respect the $G$ isometry, so that is depends only on the spectators, but this assumption can be relaxed.}
\begin{equation}
K^m=0\,,\quad K^I=n^I\,,\qquad X_m=\partial_m(\phi+\tfrac12\ln\det N)-\tilde B_{mI}n^I\,,\quad X_I=-\tilde B_{IJ}n^J\,.
\label{eq:KX}
\end{equation}
They involve the trace of the structure constants\footnote{The identification of $K$ with the trace of the structure constants was suggested earlier in \cite{Hong:2018tlp}.} of $\mathfrak g$, $n^I=\delta^{IJ}n_J$ with $n_I=f^J_{IJ}$. 
As already mentioned, in the generic case we must write the results in terms of the fields $K$ and $X$. Indeed when $n_I\neq0$ the background solves the generalized type II supergravity equations~\cite{Arutyunov:2015mqj,Wulff:2016tju} but not the standard ones, and the sigma model is scale but not Weyl invariant at one loop. When $\alg g$ is unimodular, $n_I=0$ and  we get a solution of standard type II supergravity consistent with the results of \cite{Alvarez:1994np,Elitzur:1994ri}. In that case, since $X$ is a total derivative we can write $X=d\tilde\phi$ in terms of a dual dilaton 
\be
\tilde\phi=\phi+\tfrac12\ln\det N\,.
\ee
It was shown in \cite{Wulff:2018aku} that there exist special ``trivial'' solutions of the generalized supergravity equations which solve the standard supergravity equations although $K$ is not zero. For this to happen $K$ must be null and, in addition to a condition involving the RR fields which we ignore here, it should satisfy $dK=i_KH$. Using the rules of NATD presented here the latter condition can be written as
\begin{equation}
n(\tilde G-\tilde B)=0\,.
\end{equation}
Since $(\tilde G-\tilde B)_{IJ}=N_{IJ}$ is invertible by assumption, it has no zero-eigenvector and therefore it would seem that no trivial solution can be generated by NATD. However, the condition written above is not invariant with respect to B-field gauge transformations, so that the conclusion can change. This will actually play a role in the  discussion of the closely related YB models.

\subsection{Rules of YB deformations}
For YB deformations the rules are a bit simpler in the sense that we do not have to write the background fields in the block-form as previously. The result can be phrased in different ways, see section \ref{sec:YB}. Here we will describe the results in terms of Killing vectors of the original background. The final result of our derivation is that in order to apply a YB deformation one should first construct
\be
\Theta^{\mu\nu}=k^\mu_{I}R^{IJ}k^\nu_{J}\,,
\ee
where $R^{IJ}$ solves the classical Yang-Baxter equation~\eqref{eq:CYBE} and $k^\mu_{I}$ are a collection of Killing vectors labeled by $I$ that are properly normalized so that they satisfy~\eqref{eq:comm-Kill}. Then the background metric and $B$-field of the YB model are simply obtained by the following generalization of the  open/closed string map
\begin{equation}
\tilde G-\tilde B=(G-B)[1+\eta \Theta(G-B)]^{-1}\,,
\end{equation}
where we have omitted indices $\mu,\nu$. The RR bispinor transforms as
\begin{equation}
\tilde{\mathcal S}^{12}=\hat\Lambda\mathcal S^{12}\,,\qquad \Lambda^{ab}=\eta^{ab}-2\eta E_\mu{}^a\hat N^\mu{}_\nu\Theta^{\nu\rho}E_\rho{}^b\,,
\end{equation}
where $\hat N^\nu{}_\mu=\big[\delta^{ \mu}{}_{\nu}+\eta \Theta^{\mu\rho}(G_{\rho\nu}-B_{\rho\nu})\big]^{-1}$. We further have
\be
K^{\mu}=\eta \Theta^{\mu\nu}n_{\nu}\,,\qquad
X_{\mu}=\partial_{\mu} (\phi -\frac12 \ln \det[1+\eta \Theta(G-B)])-\eta \tilde B_{\mu\nu}\Theta^{\nu\rho}n_{\rho}\,,
\ee
and, when the Killing vectors used to construct $\Theta$ define a unimodular algebra $f_{IJ}^J=0$, we find the deformed dilaton
\be
\tilde \phi = \phi -\frac12 \ln \det[1+\eta \Theta(G-B)]\,.
\ee
We refer to section~\ref{sec:YB} for the derivation and a discussion of trivial solutions for YB deformations.

\section{NATD of Green-Schwarz strings}\label{sec:NATD}
In this section we apply NATD to a generic Green-Schwarz string with isometries. To perform NATD we assume that we can bring the supervielbein to the form
\begin{equation}
E^A=(g^{-1}dg)^IE_I{}^A(z)+dz^ME_M{}^A(z)\,,\qquad\left(A=(a,\alpha)\,,\quad a=0,\ldots,9\,,\quad\alpha=1,\ldots,32\right)\,,
\label{eq:EA}
\end{equation}
with $g\in G$ encoding the coordinates we want to dualize and $z^M=(x^m,\,\theta^{\u \alpha})$ denoting the remaining (spectator) coordinates. The isometry (sub)group $G$ to be dualized acts as $g\rightarrow ug$, $z\rightarrow z$ for a constant element $u\in G$. To avoid extra awkward signs, we will present the derivation when $G$ is an ordinary Lie group, but we will write the end result for the dualized geometry such that it applies also to the case when $G$ is a super Lie group. The index $I$ takes $\dim G$ values and since we want to include the case in which a local symmetry of the sigma model (which we do not fix) is a subgroup of $G$, we allow the possibility that the total range of indices $(m,I)$ is greater than ten. In that case the local symmetry can be used at the end to remove the spurious coordinates and leave the ten physical ones. In that case $E_I{}^a$ also involves a projection matrix \cite{Lozano:2011kb}, the simplest example being a supercoset geometry where $E_I{}^a$ is proportional to the projector on the coset directions (usually denoted by $P^{(2)}$).

The (classical) Green-Schwarz string action is
\begin{equation}\label{eq:Soriginal}
S=T\int_\Sigma\,(\tfrac12E^a\wedge*E^b\eta_{ab}+B)\,,
\end{equation}
where we are using worldsheet form notation and the supervielbein $E^a$ and NSNS two-form potential $B$ are understood to be pulled back to the worldsheet $\Sigma$. To perform NATD we write this action in first order form using (\ref{eq:EA}), replacing $g^{-1}dg\rightarrow A$ and adding a Lagrange multiplier term to enforce the flatness of $A$
\begin{align}
S'=&\frac{T}{2}\int_\Sigma\,\Big(
A^I\wedge(G_{IJ}*-B_{IJ})A^J
+2dz^M\wedge(G_{MI}*-B_{MI})A^I
\nonumber\\
&\qquad{}
+(-1)^{\text{deg} N}dz^M\wedge(G_{MN}*-B_{MN})dz^N
+\nu_I(2dA^I-f^I_{JK}A^J\wedge A^K)
\Big)\,.
\label{eq:S-first}
\end{align}
The components of the (super) metric are $G_{IJ}=E_I{}^aE_J{}^b\eta_{ab}$, $G_{IM}=G_{MI}=E_I{}^aE_M{}^b\eta_{ab}$ and $G_{MN}=E_M{}^aE_N{}^b\eta_{ab}$. Integrating out $A$ gives\footnote{These solutions and the following action are written so that they hold also when $G$ is a supergroup.}
\begin{equation}\label{eq:integroutA}
(1\pm*)A^I=-(1\pm*)\left(d\nu_J+dz^M[\mp G-B]_{MJ}\right)N_\mp^{JI}\,,\qquad
N_\pm^{IJ}=\left(\pm G_{IJ}-B_{IJ}-\nu_Kf^K_{IJ}\right)^{-1}
\end{equation}
and the dual action
\begin{align}
\tilde S
=&\frac{T}{4}\int_\Sigma\,\Big\{
\left(d\nu_I+dz^M[G-B]_{MI}\right)N_+^{IJ}\wedge(1+*)\left(d\nu_J-dz^M[G+B]_{MJ}\right)
\nonumber\\
&\qquad{}
+\left(d\nu_I-dz^M[G+B]_{MI}\right)N_-^{IJ}\wedge(1-*)\left(d\nu_J+dz^M[G-B]_{MJ}\right)
\nonumber\\
&\qquad{}
+2(-1)^Ndz^M\wedge(G_{MN}*-B_{MN})dz^N
\Big\}
\nonumber\\
=&T\int_\Sigma\,\left(\tfrac12\tilde E_\pm^a\wedge*\tilde E_\pm^b\eta_{ab}+\tilde B\right)\,.
\label{eq:Sdual}
\end{align}
In the last step we have written the dualized action in Green-Schwarz form by defining two possible sets of dual supervielbeins\footnote{When $G$ is a supergroup the correct expressions are obtained by writing things in a form which is symmetric between $N_+$ and $N_-$ (and where contracted indices are adjacent), e.g.
$$
\tilde E_\pm^a=dz^ME_M{}^a
-\tfrac12\left(d\nu_I+dz^M[\mp G-B]_{MI}\right)N_\mp^{IJ}E_J{}^a
+\tfrac12E_I{}^aN_\pm^{IJ}\left(d\nu_J+dz^M[\mp G-B]_{MJ}\right)\,.
$$
This will be true also for the expressions for $\Lambda$, $K$, $X$ and $\tilde{\mathcal S}$ to be derived below.
\label{foot:super}
}
\begin{equation}
\tilde E_\pm^A=dz^ME_M{}^A-\left(d\nu_I+dz^M[\mp G-B]_{MI}\right)N_\mp^{IJ}E_J{}^A\,.
\label{eq:Etildepm}
\end{equation}
The dual B-field can also be written in two equivalent ways
\begin{equation}
\tilde B
=
\tfrac12dz^N\wedge dz^MB_{MN}
+\tfrac12\left(d\nu_I+dz^M[\pm G-B]_{MI}\right)\wedge N_\pm^{IJ}\left(d\nu_J-dz^M[\pm G+B]_{MJ}\right)\,.
\label{eq:Btilde}
\end{equation}
We choose $\tilde E_+^a$ to be the dual bosonic supervielbein, while $\tilde E_-^a$ is related to it by a Lorentz transformation as follows
\begin{equation}
\tilde E^a=\tilde E_+^a
\,,\qquad
\tilde E'^a=\tilde E_-^a=\tilde E^b\Lambda_b{}^a
\,,\qquad
\Lambda_b{}^a=\delta_b^a-2E_{Ib}N_+^{IJ}E_J{}^a\,.
\label{eq:Eatilde}
\end{equation}
This is easily seen to follow from the useful identity
\begin{equation}
\left(d\nu_I+dz^M[G-B]_{MI}\right)N_+^{IJ}
=
\left(d\nu_I-dz^M[G+B]_{MI}\right)N_-^{IJ}
+2\tilde E^a\,E_{Ia}N_+^{IJ}\,.
\label{eq:useful}
\end{equation}
It is interesting to compute the determinant of the Lorentz transformation $\Lambda$. We have (suppressing the indices)
\begin{align}
\det\Lambda
=&
\exp(\mathrm{tr}\,\ln\Lambda)
=
\exp\left(-\mathrm{tr}\sum_{n=1}^\infty\frac{(2EN_+E)^n}{n}\right)
=
\exp\left(-\mathrm{tr}\sum_{n=1}^\infty\frac{(2GN_+)^n}{n}\right)
\nonumber\\
=&
\exp[\mathrm{tr}\,\ln(1-2GN_+)]
=
\exp[\mathrm{tr}\,\ln(1-(N_+^{-1}+N_+^{-T})N_+)]
=
\det(-N_+^{-T}N_+)
\nonumber\\
=&
(-1)^{\dim G}\,.
\end{align}
This shows that this Lorentz transformation is an element of $SO(1,9)$ only when $\dim G$ is even. When $\dim G$ is odd, i.e. one dualizes on an odd number of directions, the Lorentz transformation  involves a reflection. In the latter case its action on spinors contains an odd number of gamma matrices, which means that one goes from type IIA to type IIB or vice versa, cf. (\ref{eq:Etildealpha}).

\subsection{The case with local symmetry}\label{sec:local}
Here we wish to give more details on the case when the original sigma model has a local symmetry that is a subgroup of $G$. We will explain how the results of the previous section apply also in that case.  We will assume that the action~\eqref{eq:Soriginal} is invariant under a local group $H\subset G$ that acts on $g$ from the right as\footnote{ One may equivalently discuss this local invariance by introducing a vector valued in the Lie algebra of $H$, so that integrating out such vector the original action is obtained.} $g\to gh,\ h\in H$.  Our goal will be to show that if the local $H$ invariance is not fixed before the dualization, NATD can still be applied in the usual way and the dual action naturally inherits the local $H$ symmetry. Therefore this ensures that the additional degrees of freedom can be removed also in the dual model, and that we are left only with physical ones of the correct number.

The action~\eqref{eq:Soriginal} is invariant under $g\to gh$ if the couplings are $H$ invariant and project out $\alg h$, the Lie algebra of $H$
\be
\begin{aligned}
&(\Ad_h^{-1})^K{}_I(G_{KL}*-B_{KL})(\Ad_h^{-1})^L{}_J=G_{IJ}*-B_{IJ}\,,\quad
&&y^I(G_{IJ}*-B_{IJ})=0=(G_{IJ}*-B_{IJ})y^J\,,\\
&(G_{MJ}*-B_{MJ})(\Ad_h^{-1})^J{}_I=G_{MI}*-B_{MI}\,, \qquad
&& (G_{MI}*-B_{MI})y^I=0\,.
\end{aligned}
\ee
Here $y\in \alg h$.
This local symmetry may be used to remove $\dim H$ degrees of freedom from the parametrization of $g$, so that the total number of physical bosonic fields (including spectators) is ten. We do not fix this local invariance yet, since this allows us to gauge the whole $G$ isometry and fix the gauge $g=1$ to arrive at the action~\eqref{eq:S-first}.
This first order action is still invariant under a local $H$ which is now implemented as 
\be
A\to h^{-1}Ah+h^{-1}dh,\qquad\qquad \nu \to h^{-1}\nu h\,.
\ee
Here $\nu=\nu_IT^I$ is taken to be an element of $\alg g^*$, the dual of the Lie algebra of $G$. We refer to section~\ref{sec:DTD} for our conventions regarding $\alg g^*$.
At the moment of integrating out $A^I$ from~\eqref{eq:S-first} one may worry about the invertibility of the relevant linear operators, given that the couplings project out the components in $\alg h$ as assumed above. We consider cases when the operators $\pm G_{IJ}-B_{IJ}-\nu_Kf^K_{IJ}$ are invertible on the whole algebra $\alg g$, so that also the components of $A$ in $\alg h$ can be integrated out. Obviously, since $\pm G_{IJ}-B_{IJ}$ are degenerate, the invertibility of the operators must be ensured by the term $\nu_Kf^K_{IJ}$. We recall that $\nu$ has not been gauged-fixed yet, and that we have a total of $\dim G$ such fields. In general the invertibility will hold only locally, meaning that there may be values of $\nu_I$ such that the operators $N_\pm^{IJ}$ become singular. Those loci will correspond to singularities in target space that we cannot remove. 
It is easy to check that the dual action~\eqref{eq:Sdual} is still invariant under the local $H$ symmetry, which is now simply implemented by $\nu \to h^{-1}\nu h$. We can then fix the local symmetry at the level of the dual action, at the same time making sure that we have the correct number of degrees of freedom and that the gauge fixing is done correctly.

Our reasoning is completely analogous to that of~\cite{Sfetsos:1999zm,Lozano:2011kb}. There the degenerate matrices $\pm G_{IJ}-B_{IJ}$ are regulated by taking $\pm G_{IJ}-B_{IJ}+\lambda \ (\text{Id}_{\alg h})_{IJ}$, where $\text{Id}_{\alg h}$ is the identity on $\alg h$.  The parameter $\lambda$ is kept during the dualization and sent to zero only at the end. It is clear that the $\lambda\to 0$ limit is non-singular only if the degeneracies of $\pm G_{IJ}-B_{IJ}$ are lifted by the additional term $\nu_Kf^K_{IJ}$. Therefore the way coset models are treated in~\cite{Sfetsos:1999zm,Lozano:2011kb} is analogous to ours. For concreteness we work out an explicit example in appendix~\ref{sec:example}.

\subsection{Extracting $X$ and $K$ from anomaly terms}
The easiest way to extract the generalized supergravity fields $X$ and $K$ is to look at the terms in the action induced at the quantum level by the NATD change of variables $g^{-1}dg\rightarrow A$ in the path integral measure \cite{Elitzur:1994ri}.\footnote{A more direct, but lengthier, approach uses the superspace constraints as we do below to extract the RR fields, see for example \cite{Borsato:2016ose}.} It was shown in \cite{Wulff:2018aku} that these non-local terms take the form
\begin{equation}
\tilde S_\sigma=\frac{1}{2\pi}\int_\Sigma\left(d\sigma\wedge K-d\sigma\wedge*X-\tfrac12\alpha'd\sigma\wedge*d\sigma\,|K|^2\right)\,,
\label{eq:Ssigma}
\end{equation}
where $\sigma=\partial^{-2}\sqrt gR^{(2)}$ is the conformal factor. From the first two terms it is easy to read off $X$ and $K$. To compute $\tilde S_\sigma$ we include the $\sigma$-dependent terms in the first order action (\ref{eq:S-first}). They are \cite{Elitzur:1994ri}
\begin{equation}
S_\sigma=\frac{1}{2\pi}\int_\Sigma\left(\sigma n_I d*A^I-\Phi d*d\sigma\right)\,,
\end{equation}
where $n_I=f^J_{IJ}$, the trace of the structure constants, $d*d\sigma=d^2\xi\sqrt gR^{(2)}$ and $\Phi$ is the dilaton superfield of the original background. Integrating out $A$ as before but now including these terms, and keeping track of the $\det N$ from the measure, we obtain
\begin{equation}
(1\pm*)A^I=-(1\pm*)\left(d\nu_J+dz^M[\mp G-B]_{MJ}\mp\alpha'n_Jd\sigma\right)N_\mp^{JI}
\end{equation}
and
\begin{align}
\tilde S_\sigma=&\,
\frac{1}{4\pi}\int_\Sigma\,\Big(
n_Id\sigma N_+^{IJ}\wedge(1+*)(d\nu_J-dz^M[G+B]_{MJ})
\\
&{}
-n_Id\sigma N_-^{IJ}\wedge(1-*)(d\nu_J+dz^M[G-B]_{MJ})
-2d\sigma\wedge*d(\Phi+\tfrac12\ln\det N_+)
\Big)
+\mathcal O(\alpha')\,.
\nonumber
\end{align}
Comparing to (\ref{eq:Ssigma}) we find
\begin{align}
K=&\,\tfrac12\left\{(d\nu_J+dz^M[G-B]_{MJ})N_+^{JI}-(d\nu_J-dz^M[G+B]_{MJ})N_-^{JI}\right\}n_I\,,
\label{eq:Ktilde}
\\
X=&\,d(\Phi+\tfrac12\ln\det N_+)+\tfrac12\left\{(d\nu_J+dz^M[G-B]_{MJ})N_+^{JI}+(d\nu_J-dz^M[G+B]_{MJ})N_-^{JI}\right\}n_I\,.
\label{eq:Xtilde}
\end{align}
These expressions simplify when written in terms of $\tilde G$ and $\tilde B$ as in (\ref{eq:KX}).

\subsection{Extracting the RR fields}
The simplest way to find the RR fields is to compute the superspace torsion $T^A=dE^A+E^B\wedge\Omega_B{}^A$ and compare to the superspace torsion constraints of \cite{Wulff:2013kga,Wulff:2016tju}, see e.g. \cite{Borsato:2016ose}. In particular the $E^a\wedge E^{\alpha1}$-term in $T^{\alpha2}$ takes the form\footnote{To improve the readability we suppress the spinor index $\alpha$ and drop the explicit $\wedge$'s from now on.}
\begin{equation}
T^2=-\tfrac18E^a\,(E^1\Gamma_a\mathcal S^{12})+\ldots
\label{eq:T2}
\end{equation}
from which we can read off the RR bispinor $\mathcal S$. Here $E^a$ is the bosonic supervielbein and $E^{\alpha1},\,E^{\alpha2}$ with $\alpha=1,\ldots,16$ are the two fermionic supervielbeins, corresponding to the two Majorana-Weyl spinors of type II supergravity. For convenience of the presentation we will use type IIA notation so that $E^1=\frac12(1+\Gamma_{11})E^1$ and $E^2=\frac12(1-\Gamma_{11})E^2$ but the type IIB expressions are essentially identical.

To compute $\tilde T^2$ and then extract the RR fields of the dualized model, we must first find the form of the fermionic supervielbeins $\tilde E^1,\,\tilde E^2$. We therefore start with the constraint on the bosonic torsion
\begin{equation}
T^a=-\tfrac{i}{2}E\Gamma^aE=-\tfrac{i}{2}E^1\Gamma^aE^1-\tfrac{i}{2}E^2\Gamma^aE^2\,,
\label{eq:Ta}
\end{equation}
and we can compute $\tilde T^a$ from $\tilde E^a$.\footnote{It might appear that one needs to know the spin connection to do this but this is not the case. Instead the fermionic vielbeins and spin connection can be read off by computing $d\tilde E^a$ as we will see.} By assumption the constraint on $T^a$ holds in the original model before dualization. In our adapted coordinates (\ref{eq:EA}) it takes the form\footnote{The anti-symmetrization is graded, e.g. $Y_{[M}Z_{N]}=\frac12(Y_MZ_N-(-1)^{MN}Y_NZ_M)$.}
\begin{align}
2\partial_{[M}E_{N]}{}^a+2\Omega_{[M|b|}{}^aE_{N]}{}^b=&(-1)^NiE_M\Gamma^aE_N\,,\\
\partial_ME_I{}^a+\Omega_{Mb}{}^aE_I{}^b-\Omega_{Ib}{}^aE_M{}^b=&iE_M\Gamma^aE_I\,,\\
f^K_{IJ}E_K{}^a+2\Omega_{[J|b|}{}^aE_{I]}{}^b=&iE_J\Gamma^aE_I\,.
\end{align}
We will also need the constraints on $H=dB$ which are
\begin{equation}
H=-\tfrac{i}{2}E^a\,E\Gamma_a\Gamma_{11}E+\tfrac16E^cE^bE^aH_{abc}=-\tfrac{i}{2}E^a\,E^1\Gamma_aE^1+\tfrac{i}{2}E^a\,E^2\Gamma_aE^2+\tfrac16E^cE^bE^aH_{abc}\,.
\label{eq:Hconstr}
\end{equation}
In our adapted coordinates we have
\begin{align}
&3\partial_{[M}B_{NP]}=H_{MNP}
\,,\qquad
&&2\partial_{[M}B_{N]I}=H_{MNI}
\,, \nonumber\\
&\partial_MB_{IJ}+f^K_{IJ}B_{MK}=H_{MIJ}
\,,\ 
&&3f^L_{[IJ}B_{K]L}=H_{IJK}\,,
\end{align}
where $H_{IJK}=E_K{}^CE_J{}^BE_I{}^AH_{ABC}$ etc.
Using these relations we can compute the exterior derivative of $\tilde E^a_\pm$ in (\ref{eq:Etildepm}) and we find
\begin{align}
d\tilde E_\pm^a
=&
-\tfrac{i}{2}\tilde E_\pm\Gamma^a\tilde E_\pm
+\tfrac{i}{2}\tilde E_\pm\Gamma_b(1\pm\Gamma_{11})\tilde E_\pm(\pm E_I{}^aN_\pm^{IJ}E_J{}^b)
-\tilde E_\pm^b\tilde E_\pm^C\Omega_{Cb}{}^a
\nonumber\\
&{}
\pm i\tilde E_\pm^b\,\tilde E_\pm\Gamma_b(1\mp\Gamma_{11})E_IN_\mp^{IJ}E_J{}^a
\pm\tilde E_\pm^c\tilde E_\pm^b(\Omega_{Ibc}\pm\tfrac12E_I{}^dH_{bcd})N_\mp^{IJ}E_J{}^a\,.
\label{eq:dEa}
\end{align}
Using our definition of the dualized bosonic supervielbein, $\tilde E^a=\tilde E^a_+$, this can be recast, using the definition (\ref{eq:Etildepm}), as\footnote{Note that (\ref{eq:useful}) implies $\tilde E_-=\tilde E_+-2\tilde E^aE_{Ia}N_+^{IJ}E_J$.}
\begin{equation}
\tilde T^a=d\tilde E^a+\tilde E^b\tilde\Omega_b{}^a
=
-\tfrac{i}{2}\Lambda^a{}_b\tilde E_+^1\Gamma^b\tilde E_+^1
-\tfrac{i}{2}\tilde E_-^2\Gamma^a\tilde E_-^2\,.
\end{equation}
Comparing to the standard form (\ref{eq:Ta}) we can read off the fermionic supervielbeins of the dualized model\footnote{We also find the spin connection of the dualized background
\begin{align}
\tilde\Omega^{ab}
=&\,
\tilde E_+^C\Omega_C{}^{ab}
-4i\tilde E_+^2\Gamma^{[a}E_I^2N_-^{IJ}E_J{}^{b]}
-2\tilde E^c(\Omega_{Ic}{}^{[a}-\tfrac12E_I{}^dH_{cd}{}^{[a})N_-^{IJ}E_J{}^{b]}
+\tilde E^c(\Omega_I{}^{ab}+\tfrac12E_I{}^dH_d{}^{ab})N_-^{IJ}E_{Jc}
\nonumber\\
&{}
-4i\tilde E^c\,E_I{}^{[a}N_+^{IJ}E_J^2\Gamma^{b]}E_K^2N_-^{KL}E_{Lc}
-2i\tilde E^c\,E_I{}^aN_+^{IJ}E_J^2\Gamma_cE_K^2N_-^{KL}E_L{}^b\,.
\nonumber
\end{align}
}
\begin{equation}
\tilde E^1=\hat\Lambda\tilde E_+^1\,,\qquad
\tilde E^2=\tilde E_-^2\,,
\label{eq:Etildealpha}
\end{equation}
where the action of the Lorentz transformation on spinors is defined by $\Lambda^a{}_b\Gamma^b= \hat\Lambda^T\Gamma^a\hat\Lambda$.
We are now ready to compute the fermionic torsion and extract the dualized RR fields by comparing to (\ref{eq:T2}). Following the same lines as above we find
\begin{align}
d\tilde E_-^2=&\,
\tfrac14(\Gamma_{ab}\tilde E_-^2)\,\tilde E_-^C\Omega_C{}^{ab}
+\tfrac12\tilde E_-^B\tilde E_-^AT_{AB}^2
-i\tilde E_-^1\Gamma_a\tilde E_-^1(E_I^2N_-^{IJ}E_J{}^a)
\nonumber\\
&{}
-2i\tilde E_-^a\,\tilde E_-^1\Gamma_aE_I^1(N_+^{IJ}E_J^2)
-\tilde E_-^b\tilde E_-^a(\Omega_{Iab}-\tfrac12H_{abc}E_I{}^c)N_+^{IJ}E_J^2\,.
\end{align}
Extracting the $\tilde E^a\tilde E^1$-terms we can read of the RR bispinor which takes the form
\begin{equation}
\tilde{\mathcal S}^{12}= \hat\Lambda\mathcal S^{12}+16i\hat\Lambda E_I^1N_+^{IJ}E_J^2\,.
\label{eq:S12tilde}
\end{equation}
The first term is a Lorentz transformation acting on one side of the original bispinor in agreement with the NATD transformation rules first proposed in \cite{Sfetsos:2010uq}, by  analogy with the abelian case. The second term starts at quadratic order in fermions if one dualizes on a bosonic algebra. However, in cases involving fermionic T-dualities the bosonic background is affected by the second term. In the case of a single fermionic T-duality it reproduces the transformation rule derived in \cite{Berkovits:2008ic}.\footnote{In the pure spinor formalism used there one does not directly see the Lorentz transformation acting on half of the fermionic directions since the pure spinor description has a larger symmetry with independent Lorentz transformations for bosons and the two fermionic directions. However, setting the fermions to zero $\Lambda$ becomes trivial and all transformations, including those of the RR fields, match.}

To be sure that the sigma model after NATD still has kappa symmetry, or equivalently that the background solves the generalized supergravity equations \cite{Wulff:2016tju}, one must also verify that $\tilde H=d\tilde B$ satisfies the correct constraints (\ref{eq:Hconstr}) (up to dimension zero). A direct calculation using (\ref{eq:Btilde}) and (\ref{eq:useful}) shows that $\tilde H$ is indeed of the right form (\ref{eq:Hconstr}).\footnote{One also finds
$$
\tilde H_{abc}=
-\tfrac12H_{abc}
+\tfrac32\Lambda_{[a}{}^dH_{bc]d}
-6E_{I[a}N_+^{IJ}\Omega_{|J|bc]}
-12i(E_{I[a}N_+^{IJ}E_{|J|}^2)\Gamma_b(E_{|K|c]}N_+^{KL}E_L^2)\,.
$$
} This proves that the dual model is indeed a Green-Schwarz string invariant under the standard kappa symmetry transformations, and it completes the derivation of the dualized target space fields which therefore solve the equations of (generalized) supergravity \cite{Wulff:2016tju}.

\section{Deformations}\label{sec:def}
NATD may be viewed as a solution-generating technique for supergravity backgrounds. Here we slightly modify the procedure to generate \emph{continuous} deformations of the dual model, which will be called deformed T-dual (DTD) models. Later we will show that a subclass of DTD models may be recast in the form of a deformation that reduces to the original sigma model when sending the deformation parameter to zero. This subclass will be identified with a generalization of YB deformations.

\subsection{Deformed T-dual models}\label{sec:DTD}
In order to define DTD models, we start from the original sigma-model, before applying NATD, and we shift the $B$-field as
\be
B_{IJ} \to B_{IJ}-\zeta\ \omega_{IJ}\,.
\ee
Here $\omega_{IJ}$ is constant and anti-symmetric in its indices. We use $\zeta$ as a parameter to keep track of the shift, or in other words the deformation. The shift affects only the components of the $B$-field along $\alg g$, and it does not spoil the global $G$ isometry.  We demand that the new term appearing in the action (i.e. $\zeta(g^{-1}dg)^I\wedge\omega_{IJ}(g^{-1}dg)^J$) should not modify the theory \emph{on-shell}, in other words that it should be a closed $B$-field. It is easy to see that this happens if and only if $\omega_{IJ}$ satisfies the 2-cocycle condition
\be\label{eq:2cc-comp}
\omega_{I[J}f_{KL]}^I
=0\,,
\ee
where the antisymmetrization involves all three indices $J,K,L$. We further demand that the $B$-field $\zeta(g^{-1}dg)^I\wedge\omega_{IJ}(g^{-1}dg)^J$ is closed but not exact, i.e. the shift should not be a gauge transformation. Thanks to this additional condition, after applying NATD the resulting deformation is non-trivial, i.e. the $\zeta$-dependence cannot be removed by a field redefinition. The non-exactness of $B$ is equivalent to $\omega_{IJ}$ not being a coboundary, i.e. $\omega_{IJ}\neq c_Kf_{IJ}^K$ for any constant vector $c_K$. Non-trivial deformations are therefore classified by elements of the second Lie algebra cohomology group $H^2(\mathfrak g)$.

\vspace{12pt}

We can view the 2-cocycle as an element of $\alg g^*\otimes \alg g^*$ by writing $\omega =  \omega_{IJ} T^I \wedge T^J$. Alternatively we may view it as a map from $\mathfrak g$ to the dual vector space (we continue to call this $\omega$ without fear of creating confusion) $\omega: \alg g \to \alg g^*$, whose action is given by 
\be
\omega(T_K)= \omega_{IJ} T^I \tr(T^JT_K)=\omega_{IK}T^I\,.
\ee
To proceed further we will endow the dual vector space $\alg g^*$ with a Lie algebra structure with structure constants $\tilde f^{IJ}_K$ so that $\alg g$ has a bialgebra structure. Therefore $\alg g\oplus\alg g^*$ becomes a Lie algebra with Drinfel'd double commutation relations\footnote{This is very similar to how one realizes NATD as a special case of Poisson-Lie T-duality \cite{Klimcik:1995ux} and it would be interesting to consider the extension of our construction to the Poisson-Lie case.}
\be
[T_I,T_J]=f_{IJ}^KT_K\,,\qquad
[T^I,T^J]=\tilde f^{IJ}_KT^K\,,\qquad
[T_I,T^J]=f_{KI}^JT^K+\tilde f^{JK}_IT_K\,.
\ee
This is always possible since we can always take $\alg g^*$ to be abelian with $\tilde f^{IJ}_K=0$. In general this construction is far from unique and there exist many possible choices of Lie algebra structure on $\alg g^*$, however this choice will have no effect in what follows.
The 2-cocycle condition \eqref{eq:2cc-comp} can now be written
\be\label{eq:2cc-op}
\omega[T_I,T_J]=P^T([\omega T_I,T_J]+[T_I,\omega T_J])\,,
\ee
where $P^T$ projects on $\alg g^*$. Note that if we take $\alg g^*$ to be abelian we can drop the projector and this equation just says that $\omega$ is a derivation on the Lie algebra $\alg g\oplus\alg g^*$. This is the choice that is most useful for the general discussion here.\footnote{In the PCM case considered in~\cite{Borsato:2016pas} or the supercoset model case considered in~\cite{Borsato:2017qsx} there is a natural Lie algebra structure on $\alg g^*$, inherited from the full isometry group. This is the structure that was chosen in~\cite{Borsato:2016pas,Borsato:2017qsx}. Nevertheless, as already mentioned this choice has no consequence in our construction, and a more natural choice may be for example to take $\alg g^*$ abelian.}

\vspace{12pt}

Apart from the shift $B_{IJ}\rightarrow B_{IJ}-\zeta \omega_{IJ}$, nothing changes in the derivation of the transformation of the action and of the background fields under NATD. Therefore, the transformation rules derived in section~\ref{sec:NATD} and presented in section~\ref{sec:summary-NATD} are valid also for DTD if we shift $B_{IJ}\rightarrow B_{IJ}-\zeta \omega_{IJ}$. The resulting DTD background is a deformation of the NATD background, and it reduces to it when $\zeta=0$.
We refer to~\cite{Borsato:2016pas,Borsato:2017qsx} for some explicit examples of DTD models obtained from PCM or from the superstring on $AdS_5\times S^5$.

\subsection{Yang-Baxter deformations}\label{sec:YB}
We will now construct deformations of the original background, rather than its NATD. We introduce a deformation parameter $\eta$ such that $\eta=0$ gives back the original sigma model. These deformations will be  obtained from the DTD construction, where we identify $\eta=\zeta^{-1}$. We identify them with Yang-Baxter deformations, since they are generated by solutions of the classical Yang-Baxter equation and they generalize the original construction for PCM and (super)cosets to generic (Green-Schwarz) sigma models. 

 The construction is possible when $\omega_{IJ}$ is invertible. Writing $R=\omega^{-1}:\alg g^*\to \alg g$ it is easy to verify that the 2-cocycle condition for $\omega$ implies that $R$ solves the classical Yang-Baxter equation
\be\label{eq:CYBE}
[Rx,Ry]-R([Rx,y]+[x,Ry])=0\,,\qquad \forall x,y\in \alg g^*,\quad\iff\quad
R^{L[I}R^{|M|J}f_{LM}^{K]}=0\,,
\ee
where the action of the operator is again defined by $R(T^I)=T_KR^{KI}$. The above is equivalent to the more familiar form of the classical Yang-Baxter equation
\be
[r_{12},r_{13}]+[r_{12},r_{23}]+[r_{13},r_{23}]=0\,,
\ee
written in terms of $r=R^{IJ}T_I\wedge T_J\in \alg g \otimes \alg g$, where the subscripts of $r_{ij}$ denote the spaces in $\alg g\otimes \alg g\otimes \alg g$ where it acts.
To recast the DTD model as a deformation of the original model we need to replace the coordinates $\nu_I$, which parametrize the dual space, by a group element $g\in G$. The invertible map $\omega: \alg g\rightarrow\alg g^*$ allows us to do this by writing \cite{Borsato:2016pas}
\be
\nu_I = \zeta\tr\left(T_I \frac{1-\Ad_g^{-1}}{\log \Ad_g}\omega \log g\right)\,.
\ee
Using the 2-cocycle condition it can be shown that this implies\footnote{The easiest way to show this is to extend $\omega$ to act as a derivation on the universal enveloping algebra of $\alg g$. With this definition we can write $\eta\nu=g^{-1}\omega(g)\in\mathfrak g^*$. We can now compute $d\nu$ and the two equivalent expressions $\omega(dg)=\omega(gg^{-1}dg)=\eta g\nu g^{-1}dg+g\omega(g^{-1}dg)$ and $\omega(dg)=\omega(dgg^{-1}g)=\omega(dgg^{-1})g+\eta dg\nu$. This gives us the two equations.
}
\be
d\nu_I=\eta^{-1}\left(R_g^{-1}(g^{-1}dg)\right)_I\,,\qquad
\nu_Kf_{IJ}^K=\eta^{-1}R^{-1}_{IJ}-\eta^{-1}(R_g^{-1})_{IJ}\,,
\label{eq:dnu}
\ee
where $R_g=\Ad_g^{-1}R\Ad_{g}$. Using this in the definition of $N^{IJ}$ in (\ref{eq:NIJ}) we get
\be
N=\eta R_g\left(1+\eta(G-B)R_g\right)^{-1}=\eta \left(1+\eta R_g(G-B)\right)^{-1}R_g\,.
\ee
With these substitution rules it is easy to check that the DTD action is recast into the following form\footnote{We still use tilde to denote transformed metric and $B$-field, but now they differ from the ones of NATD. The transformations rules are given below.} 
\begin{align}
S=&\frac{T}{2}\int_\Sigma\,\Big(
(g^{-1}dg)^I\wedge(\tilde G_{IJ}*-\tilde B_{IJ})(g^{-1}dg)^J
+2dz^M\wedge(\tilde G_{MI}*-\tilde B_{MI})(g^{-1}dg)^J
\nonumber\\
&\qquad{}
+(-1)^Ndz^M\wedge(\tilde G_{MN}*-\tilde B_{MN})dz^N
-\eta^{-1} (dgg^{-1})^I\wedge \omega_{IJ}(dgg^{-1})^J
\Big)\,,
\label{eq:SYB}
\end{align}
where we isolated the last term which does not behave well in the $\eta\to 0$ limit. This term is again a closed $B$-field thanks to the 2-cocycle condition satisfied by $\omega$, and therefore it does not contribute to the equations of motion. We define the action of the YB model as the above one where the closed $B=\eta^{-1} (dgg^{-1})^I\wedge \omega_{IJ}(dgg^{-1})^J$ is removed. Dropping it we do not modify the on-shell theory, so that if the original model is classically integrable this property is inherited also by the YB deformation. In this way we can also achieve a non-singular $\eta\to 0$ limit, which yields the original undeformed model as is clear from the expressions given below.
This also implies that YB deformations may be viewed as interpolations between the original model (obtained just by sending $\eta\to 0$) and the dual one (which is recovered in the equivalent DTD formulation after sending $\zeta\to 0$, which is $\eta\to \infty$).

Setting fermions to zero and assuming a bosonic group $G$, we then read off
\begin{align}
&\tilde G_{mn}=G_{mn}-\eta\big[(G-B)\hat NR_g(G-B)\big]_{(mn)}\,,\label{eq:GYB}\\
&\tilde G_{mI}=\tfrac12\big[(G-B)\hat N\big]_{mI}+\tfrac12\big[\check N(G-B)\big]_{Im}\,,\qquad
\tilde G_{IJ}=\big[(G-B)\hat N\big]_{(IJ)}\,,\nonumber\\
\nonumber\\
&\tilde B_{mn}=B_{mn}+\eta\big[(G-B)\hat NR_g(G-B)\big]_{[mn]}\,,\label{eq:BYB}\\
&\tilde B_{mI}=-\tfrac12\big[(G-B)\hat N\big]_{mI}+\tfrac12\big[\check N(G-B)\big]_{Im}\,,\qquad
\tilde B_{IJ}=-\big[(G-B)\hat N\big]_{[IJ]}\,,\nonumber
\end{align}
while the RR bispinor is again transformed by a Lorentz transformation $\hat\Lambda$ acting on spinor indices from the left\footnote{For YB deformations $\Lambda\in SO(1,9)$ and it is therefore useful to parametrize it in terms of an anti-symmetric matrix $A^{ab}$ as $\Lambda=(1+A)^{-1}(1-A)$ which implies $A = (1-\Lambda)(1+\Lambda)^{-1}$, where we lowered one index with $\eta_{ab}$ to obtain e.g. $\Lambda^a{}_b$.
Then the Lorentz transformation on spinor indices  $\Lambda_a{}^b \Gamma_b= \hat\Lambda^T \Gamma_a\hat\Lambda$ can be written as a finite sum~\cite{Hassan:1999mm}
\be
\hat\Lambda=[\det (\eta+A)]^{-1/2} \AE(-\tfrac{1}{2}A_{ab}\Gamma^{ab}),\qquad
\AE(\tfrac{1}{2}A_{ab}\Gamma^{ab})\equiv 1+\sum_{n=1}^{n=5}\frac{1}{n!2^n}A_{a_1b_1}\cdots A_{a_nb_n}\Gamma^{a_1b_1\cdots a_nb_n}\,.
\ee}
\begin{equation}
\tilde{\mathcal S}^{12}= \hat\Lambda\mathcal S^{12}\,,\qquad \Lambda^{ab}=\eta^{ab}-2\eta E_I{}^a\hat N^I{}_J(R_g)^{JK}E_K{}^b\,.
\end{equation}
In the above we have also defined
\begin{equation}
\hat N^J{}_I=\big[\delta^{ I}{}_{J}+\eta (R_g)^{IK}(G_{KJ}-B_{KJ})\big]^{-1}\,,\quad 
\check N_I{}^J=\big[\delta_{J}{}^{ I}+\eta (G_{JK}-B_{JK}) (R_g)^{KI}\big]^{-1}=\big[R_g^{-1}\hat NR_g\big]_I{}^J\,.
\end{equation}
Using (\ref{eq:dnu}) in (\ref{eq:Ktilde}) and (\ref{eq:Xtilde}) we find\footnote{In the expression for $X$ we have used the fact that $d(\ln\det[\eta R_g])=\tr(R_g^{-1}dR_g)=2f^I_{JI}[g^{-1}dg]^J=2(g^{-1}dg)^In_I$.}
\begin{align}
K^m=0\,,\quad K^I=\eta[R_gn]^I\,,\quad
X_m=\partial_m(\phi+\tfrac12\ln\det\hat N)-\eta\tilde B_{mI}[R_gn]^I\,,\quad
X_I=-\eta\tilde B_{IJ}[R_gn]^J\,.
\end{align}
At this point we wish to comment on the possibility of having ``trivial'' solutions of the generalized supergravity equations, namely ones that solve the more restricting standard supergravity equations while $K$ does not vanish. This is possible if~\cite{Wulff:2018aku} 
\begin{equation}
0=K^I(\tilde G-\tilde B)_{IJ}=
-\eta[nR_g(G-B)\hat N]_J
=
[n(\hat N-1)]_J
\quad\iff
\quad K^I(G-B)_{IJ}=0\,,
\end{equation}
i.e. the \emph{original} $G-B$ must be degenerate. Such trivial solutions are possible for YB deformations since we do not need to assume that $G-B$ is non-degenerate. They are, at least naively, not possible for NATD since there they would imply that the \emph{dual} $\tilde G-\tilde B$ is degenerate, which is not allowed by assumption, see section~\ref{sec:NATD}. This discrepancy has to do with the fact that when going from DTD to YB we did not just change coordinates, we also shifted $B$ by dropping the extra closed term in~\eqref{eq:SYB}. Explicit trivial solutions were found in~\cite{Hoare:2016hwh}, and more recently in~\cite{Sakamoto:2018krs} by double field theory $\beta$-shifts starting from $AdS_3\times S^3\times T^4$ with non-zero $B$-field. It is clear from the present discussion that these solutions can be equivalently generated from the construction of YB deformations provided here. An example is provided in section \ref{sec:AdS3-ex}.

\subsubsection{A convenient rewriting}
As remarked in the introduction the deformed metric and B-field can be obtained from the original $G$ and $B$ by the following generalization of the open/closed string map used by Seiberg and Witten
\begin{equation}\label{eq:SWM}
\tilde G-\tilde B=(G-B)[1+\eta R_g(G-B)]^{-1}\,.
\end{equation}
This is readily seen after noticing that, since $R_g$ has only $IJ$ indices, the following operator is of block form
\be\label{eq:block}
1+\eta R_g(G-B)
=\left(
\begin{array}{cc}
\delta^m{}_n & 0\\
\eta [R_g(G-B)]^I{}_n & \delta^{ I}{}_{J}+\eta [R_g(G-B)]^{ I}{}_{J}
\end{array}
\right)\,,
\ee
and it is straightforward to invert it giving
\be
[1+\eta R_g(G-B)]^{-1}
=
\left(
\begin{array}{cc}
\delta^m{}_n & 0\\
-\eta [\hat NR_g(G-B)]^I{}_n & \hat N^I{}_J
\end{array}
\right)\,,
\ee
where we used $\hat N^J{}_I=[\delta^{ I}{}_{J}+\eta (R_g)^{IK}(G_{KJ}-B_{KJ})]^{-1}$. It is easy to check that~\eqref{eq:SWM} indeed reproduces the formulas (\ref{eq:GYB}--\ref{eq:BYB}) for the transformed metric and $B$-field.

So far we have worked with explicit group elements and algebra indices. It is sometimes convenient to translate the results so that the information on the initial isometries of the model is encoded in a set of Killing vectors. Thanks to this rewriting the YB deformation may be applied without the need of introducing an explicit parametrization of the group $G$.
Isometries of the metric and $B$-field are translated into equations for a family of Killing vectors $k_{I}^{\mu}$, where $I=1,\ldots,\dim(G)$ is the index to enumerate them. In particular, the metric possesses an isometry when shifting infinitesimally the coordinates  $X^{\mu} \to X^{\mu}+\epsilon^{I}k_{I}^{\mu}+\mathcal{O}(\epsilon^2)$, if $k_{I}^{\mu}$ satisfy the Killing vector equation
\be
\nabla_{\mu}k_{I\, \nu}+\nabla_{\nu}k_{I\, \mu}=0\,.
\ee
In order to make a connection with the formulation in terms of the group element $g$, it is enough to notice that its variation $\delta g$ under an infinitesimal transformation  can be understood in two ways, either as $\delta x^i\partial_i g$, or as $\epsilon^{I}T_{I}g$, the latter being the infinitesimal version of the global transformation $g\to \exp(\epsilon^{I}T_{I})g$. We recall that indices $i,j$ are used to label coordinates $x^i$ on the group $G$. This leads to the identification
\be
k_{I}^J\equiv k_{I}^\mu\ell_\mu^J=\tr(T^J\Ad_g^{-1}T_{I})=(\Ad_g^{-1})_{I}^J\,,\quad \text{where } g^{-1}dg=\ell^IT_I\,.
\ee
Obviously, $\ell_{\mu}^I$ and $k_{I}^{\mu}$ are non-zero only for $\mu=i$. The structure constants of the Lie algebra may be recovered by computing
\be\label{eq:comm-Kill}
\mathcal L_{k_I}k_{J\mu}-\mathcal L_{k_J}k_{I\mu}=-f_{IJ}^Kk_{K\mu}\,,
\ee
where $\mathcal L$ is the Lie derivative. Now let us notice that we can rewrite
\be\label{eq:Theta}
\Theta^{IJ}\equiv(R_g)^{IJ}=k^I_{K}R^{KL}k^J_{L}\,,
\ee
and that, before fixing any local symmetry (if present), the matrix $\ell_i^I$ is invertible. Let us denote the inverse by $\ell^i_I$ so that $\ell_i^I\ell^i_J=\delta^I_J$. This allows us to convert all algebra indices $I,J$ in~\eqref{eq:SWM} into curved indices $i,j$. Therefore the YB deformation of the metric and $B$-field may also be written as
\begin{equation}\label{eq:SWMT}
\tilde G-\tilde B=(G-B)[1+\eta \Theta(G-B)]^{-1}\,.
\end{equation}
This formula is then equally valid both when we use indices $\{m,I\}$ or $\{m,i\}$.
When a local symmetry is present we arrive at the same result since the local invariance can be left unfixed until the end.
With a similar reasoning we may rewrite also the transformation rule for the dilaton when $n_I=f_{IJ}^J=0$. In fact, when computing the determinant of $\hat N^I{}_J$ we may as well extend it to all $\mu,\nu$ indices. Since the (inverse of the) operator is in the block-form~\eqref{eq:block}, it is clear that $\det(\hat N^{\mu}{}_{\nu})=\det(\hat N^I{}_J)$. This also means that we can obtain the deformed dilaton simply by calculating
\be\label{eq:dilaton-T}
\tilde \phi = \phi -\frac12 \ln \det[1+\eta \Theta(G-B)]\,.
\ee
More generally, when $n_I\neq 0$ we may write
\be
K^{\mu}=\eta \Theta^{\mu\nu}n_{\nu}\,,\qquad
X_{\mu}=\partial_{\mu} \tilde \phi-\eta \tilde B_{\mu\nu}\Theta^{\nu\rho}n_{\rho}\,.
\ee


\subsection{Two examples of YB deformations}
We wish to work out two examples of YB deformations that do not fall under the (super)coset construction. In addition to the intrinsic interest of the following (deformed) backgrounds, the calculations also illustrate the applicability of our method.

\subsubsection{YB deformation of the D3-brane background}\label{sec:brane}
Our first motivation is to understand a YB deformation of $AdS_5\times S^5$ generated by an $R$-matrix that cannot be interpreted as a sequence of TsT transformations. In particular, we want to use the formula~\eqref{eq:SWMT} to ``uplift'' the YB deformation from the $AdS_5\times S^5$ background to the full D3-brane background, before taking the near-horizon limit. This is in the spirit of~\cite{vanTongeren:2015uha,vanTongeren:2016eeb}, where the uplift to the brane background was done for YB deformations that are (sequences of) TsT transformations.
For the sake of the discussion we focus on the NS-NS sector, where the dilaton is constant (we set it to zero for simplicity), $B=0$ and the metric is
\be\label{eq:metric-3brane}
ds^2=H^{-1/2}\, dx_i dx^i+H^{1/2}(dr^2+r^2 \, ds_{S^5}^2)\,, \quad\qquad H=1+\frac{(\alpha')^2L^4}{r^4}\,,
\ee
where $i=0,\ldots,3$ and $\eta_{ij}=\text{diag}(-1,1,1,1)$. The above metric has an $ISO(1,3)$ Poincar\'e isometry acting on the $x^i$ coordinates, and an $SO(6)$ isometry acting on the five-dimensional sphere $S^5$. We will now deform the background by exploiting the Poincar\'e part of the isometries. The Killing vectors in this case may be written as
\be
\text{Translations: } k_{[p_i]}^\mu=\delta^\mu_i\,,\qquad
\text{Lorentz: } k_{[J_{ij}]}^\mu=-\delta^\mu_ix_j+\delta^\mu_jx_i\,,\qquad
i,j=0,\ldots,3.
\ee
We wish to ``uplift'' the YB deformation of $AdS_5\times S^5$ worked out in section 6.4 of~\cite{Borsato:2016ose}, where the $R$-matrix was chosen to be 
\begin{equation}
R=p_1\wedge p_3 + (p_0+p_1)\wedge (J_{03}+J_{13})\,.
\end{equation}
That is possible since this $R$-matrix is constructed out of generators that are isometries also of the D3-brane background before the near-horizon limit. Following~\eqref{eq:Theta} we therefore construct
\be\label{eq:Theta-n4}
\Theta^{\mu\nu}=2\left[k_{[p_1]}^\mu k_{[p_3]}^\nu+(k_{[p_0]}^\mu+k_{[p_1]}^\mu)(k_{[J_{03}]}^\nu+k_{[J_{13}]}^\nu)\right]-\mu\leftrightarrow \nu\,.
\ee
More explicitly, in the block with $\mu,\nu=0,\ldots,3$ it is
\be
\Theta^{\mu\nu}=2\left(
\begin{array}{cccc}
 0 & 0 & 0 & -x^- \\
 0 & 0 & 0 & -x^-+1 \\
 0 & 0 & 0 & 0 \\
x^- & x^--1 & 0 & 0 \\
\end{array}
\right)\,,
\ee
where we introduced the standard light-cone coordinates $x^\pm=x^0\pm x^1$. Now, using~\eqref{eq:SWMT} and~\eqref{eq:dilaton-T} we obtain the following deformed metric, $B$-field and dilaton
\be
\begin{aligned}
d\tilde s^2=&
-\frac{\hat\eta ^2 \xi_-^2H^{-1/2} d\xi_-^2}{4    \left(H-4 \hat\eta ^2 \xi_-\right)}
-\frac{H^{-1/2} \left(H-2 \hat\eta ^2 \xi_-\right)d\xi_- dx^+}{2  \left(H-4 \hat\eta ^2 \xi_-\right)}
-\frac{\hat\eta ^2   H^{-1/2}(dx^+)^2}{ \left(H-4 \hat\eta ^2 \xi_-\right)}\\
&+H^{-1/2}dx_2^2
+\frac{H^{1/2}   dx_3^2}{H-4 \hat\eta ^2 \xi_-}
+H^{1/2}  (dr^2+r^2 \, ds_{S^5}^2)\,,\\
\tilde B=&\frac{\hat\eta}{2}\frac{dx^3\wedge(2dx^++\xi_-d\xi_-)}{H-4\hat\eta^2\xi_-}\,,\qquad\qquad
\exp{(-2\tilde \phi)}=1-\frac{4\hat\eta^2\xi_-}{H}\,.
\end{aligned}
\ee
We chose $\hat\eta$ as deformation parameter and to simplify expressions we redefined $\xi_-= 2x^--1$. We now want to check that the near-horizon geometry of this YB deformation of the D3-brane background indeed yields the YB deformation of $AdS_5\times S^5$ of~\cite{Borsato:2016ose}. In the near-horizon limit one sends $r\to 0$ and $\alpha'\to 0$ while keeping the ratio $r/\alpha'$ fixed. We achieve this by rewriting $r=\alpha'L^2 /z$ and $\hat\eta=\eta L^{-2}/\alpha'$, and then sending $\alpha'\to 0$. We obtain
\be
\begin{aligned}
\lim_{\alpha'\to 0}\frac{ds^2}{\alpha'L^2}=&
z^{-6}\left(1-\frac{4 \eta   ^2 \xi_-}{z^4}\right)^{-1}\left[z^4 d {x_3}^2-\eta ^2 (d {x^+})^2-\frac{1}{4} d\xi_-  \left(\eta ^2 \xi_- ^2 d\xi_- +2 d {x^+} \left(z^4-2 \eta ^2 \xi_-
   \right)\right)\right]\\
&+\frac{d {x_2}^2+dz^2}{z^2}+ds_{S^5}^2\\
\lim_{\alpha'\to 0}\frac{B}{\alpha'L^2}=&\frac{\eta}{2}\frac{dx^3\wedge(2dx^++\xi_-d\xi_-)}{z^4-4\eta^2\xi_-},\qquad\qquad
\lim_{\alpha'\to 0}e^{-2\phi}=1-\frac{4\eta^2\xi_-}{z^4}\,,
\end{aligned}
\ee
which indeed reproduces\footnote{In this paper we have a different convention for the sign of the $B$-field.} (the NS-NS sector of) the deformation of $AdS_5\times S^5$ appearing in section 6.4 of~\cite{Borsato:2016ose}.
Uplifting the YB deformation to the D3-brane background is particularly interesting since it also allows us to go far from the brane and understand how the flat space in which it is embedded has been deformed. In the limit $r\to \infty$ we have simply $H\to 1$
\be
\begin{aligned}
ds^2=&
-\frac{\hat\eta ^2 \xi_-^2 d\xi_-^2}{4    \left(1-4 \hat\eta ^2 \xi_-\right)}
-\frac{ \left(1-2 \hat\eta ^2 \xi_-\right)d\xi_- dx^+}{2  \left(1-4 \hat\eta ^2 \xi_-\right)}
-\frac{\hat\eta ^2  (dx^+)^2}{ \left(1-4 \hat\eta ^2 \xi_-\right)}\\
&+dx_2^2
+\frac{ dx_3^2}{1-4 \hat\eta ^2 \xi_-}
+ ds_{R^6}^2\\
B=&\frac{\hat\eta}{2}\frac{dx^3\wedge(2dx^++\xi_-d\xi_-)}{1-4\hat\eta^2\xi_-},\qquad\qquad
e^{-2\phi}=1-4\hat\eta^2\xi_-\,.
\end{aligned}
\ee
Obviously, the above background may be also obtained directly as a YB deformation of flat space with $\Theta$ given by~\eqref{eq:Theta-n4}. In the AdS/CFT correspondence one looks at open strings stretching between D3-branes in flat space, whose low-energy limit produces $\mathcal N=4$ super Yang-Mills.
In the presence of a $B$-field as in the case considered here, open strings feel an effective metric $g_{\mu\nu}$ and a non-commutativity parameter $\theta^{\mu\nu}$ that are related to the metric and $B$-field $G_{\mu\nu},B_{\mu\nu}$ of the closed string by\footnote{As it is written, this open/closed string map assumes the invertibility of $(G-B)$. The generalization (of the inverse transformation) to the case of degenerate $(G-B)$ is in fact given by our~\eqref{eq:SWMT}.} \cite{Seiberg:1999vs}
\be
g^{\mu\nu}+\frac{\theta^{\mu\nu}}{2\pi\alpha'}=(G_{\mu\nu}-B_{\mu\nu})^{-1}\,,
\ee
where $g^{\mu\nu}$ is obviously obtained by taking the symmetric part of the right-hand-side, while $\theta^{\mu\nu}$ the antisymmetric part. In general, if we apply the open/closed string map to a background obtained by a YB deformation we get
\be\begin{aligned}
&g^{-1}+\frac{\theta}{2\pi\alpha'}=(\tilde G-\tilde B)^{-1}=[(G-B)^{-1}+\eta \Theta]\,,\\
\implies &g^{-1}=(G-B)^{-1}_s\,,\qquad
\theta=2\pi\alpha'[(G-B)^{-1}_a+\eta\Theta]\,,
\end{aligned}
\ee
where we directly relate the open-string quantities to the metric and $B$-field  $G,B$ of the original model before the YB deformation, and subscripts $s$ and $a$ indicate the symmetric and antisymmetric parts. In our specific example, before deforming, the brane system is in a flat spacetime with vanishing $B$-field, meaning that the effective open-string metric will coincide with the flat one, and the non-commutativity parameter will be essentially defined by the YB $R$-matrix
\be
g_{\mu\nu}=G_{\mu\nu}\,,\qquad
\theta^{\mu\nu}=2\pi\alpha'\hat\eta\, \Theta^{\mu\nu}\,.
\ee
This discussion is obviously generic and is not confined to the current example.
Apart from uncovering the non-commutativity structure, at this point one should also take the low-energy limit of open strings in the non-commutative spacetime. Here we are considering a case with an electric $B$-field, and these instances are known to produce problems when trying to take the low-energy limit~\cite{Seiberg:2000ms}. It is therefore not clear whether the low-energy limit yields a non-commutative gauge theory with $\theta$ as non-commutativity parameter.
The relation between gravity duals of non-commutative gauge theories and YB deformations was first pointed out in~\cite{Matsumoto:2014gwa}.

Certain YB deformations of $AdS_5\times S^5$ are constructed out of generators that are not isometries of the brane background and that become isometries only after taking the near-horizon limit. For these examples it is not clear how to uplift the YB deformation to the brane background. It would be interesting to see if YB deformations can be extended also to cases without isometries by using Poisson-Lie T-duality.

\subsubsection{YB deformation of $AdS_3\times S^3\times T^4$ with $H$-flux}\label{sec:AdS3-ex}
We now want to apply the YB deformation to a background with degenerate $G-B$, and we will compare our results to those of~\cite{Sakamoto:2018krs}. There it was indeed shown that YB deformations of $AdS_5\times S^5$ are equivalent to local $\beta$-transformations of the double theory, and it was proposed that local $\beta$-shifts should be the natural way to generalize YB deformations to generic backgrounds, including cases with degenerate $G-B$.
The example we consider is that of $AdS_3\times S^3\times T^4$ with non-vanishing $H$-flux
\be
\begin{aligned}
ds^2&=\frac{dx_i dx^i+dz^2}{z^2}+ds_{S^3}^2+ds_{T^4}^2\,,
\quad ds_{S^3}^2=\frac{1}{4}\left[d\theta^2+\sin^2\theta d\varphi^2+(d\psi+\cos \theta d\varphi)^2\right]\\
B&=\frac{dx^0\wedge dx^1}{z^2}+\frac{1}{4}\cos \theta d\varphi \wedge d\psi\,.
\end{aligned}
\ee
$G-B$ is degenerate because of the rows (or columns) $i=0,1$. The dilaton is constant and for simplicity we set it to zero. To generate a YB deformation we will make use of the Killing vectors of the Poincar\'e isometry
\be
\text{Translations: } k_{[p_i]}^\mu=\delta^\mu_i\,,\qquad
\text{Lorentz: } k_{[J_{ij}]}^\mu=-\delta^\mu_ix_j+\delta^\mu_jx_i\,,\qquad
i,j=0,1\,.
\ee
In order to compare to the results of section 4.2.2 of~\cite{Sakamoto:2018krs} we take  $R=c^i p_i\wedge J_{01}$ or
\be
\Theta^{\mu\nu}=(c^i k_{[p_i]}^\mu) k_{[J_{01}]}^\nu-\mu\leftrightarrow \nu\,, 
\ee
where we sum over $i=0,1$. The classical YB equation is satisfied only when the parameters satisfy $c^0=\pm c^1$. Now using~\eqref{eq:SWMT} and~\eqref{eq:dilaton-T} we obtain the YB deformed background
\be
\begin{aligned}
ds^2&=\frac{dx_i dx^i}{z^2-2\eta c_j x^j}+\frac{dz^2}{z^2}+ds_{S^3}^2+ds_{T^4}^2\,,\\
B&=\frac{dx^0\wedge dx^1}{z^2-2\eta c_j x^j}+\frac{1}{4}\cos \theta d\varphi \wedge d\psi\,,
\qquad e^{-2\phi}=1-\frac{2\eta c_i x^i}{z^2}\,,
\end{aligned}
\ee
which agrees with the background obtained in section 4.2.2 of~\cite{Sakamoto:2018krs}. This confirms in a specific example the expected equivalence of YB deformations and local $\beta$-shifts even beyond the standard ($H=0$) supercoset case. As already noticed in~\cite{Sakamoto:2018krs} the above background is actually a trivial solution since the vector $K$ decouples from the generalized supergravity equations.

\section{Conclusions}\label{sec:conclusions}
We have derived the transformation rules for the supergravity fields under NATD by carrying out the dualization in the general case for the Green-Schwarz string. This generalizes the derivation performed for the case of the supercoset in~\cite{Borsato:2017qsx}. If the dualized group $G$ is not unimodular there is in general an anomaly, which is reflected in the fact that the resulting background solves the generalized supergravity equations of \cite{Arutyunov:2015mqj,Wulff:2016tju} rather than the standard ones. We have also discussed a generalization where one adds a closed $B$-field to the action prior to performing the duality transformation. This leads to so-called DTD models and, in special cases, a generalization of Yang-Baxter models \cite{Klimcik:2002zj,Klimcik:2008eq}. We have also seen that this gives us an interesting way to find examples that avoid the anomaly from non-unimodularity of $G$ along the lines discussed in \cite{Wulff:2018aku}.

Non-abelian T-duality can be embedded in the even more general framework of Poisson-Lie T-duality \cite{Klimcik:1995ux}. Also this case can be formulated at the path integral level and an anomaly arises in a similar way \cite{Tyurin:1995bu} (see also \cite{VonUnge:2002xjf}). It would be interesting to extend our analysis to this case which would also make further contact with \cite{Lust:2018jsx}. It would also allow us to extend DTD and YB deformations to cases without isometries, and perhaps help to uplift all YB deformations of $AdS_5\times S^5$ to deformations of the brane background. It would also be interesting to consider the case of open strings along the lines of the recent paper \cite{Cordonier-Tello:2018zdw}.

We have found that a natural way to rephrase YB deformations is in terms of a generalization of (the inverse of) the open/closed string map of Seiberg and Witten, thus extending what was observed in the case of both homogeneous and inhomogeneous YB deformations of PCM or (super)cosets. Since the inhomogeneous case cannot be formulated in terms of our construction we have only considered the homogeneous one here, but it would be interesting to see what happens if we take $R$ in~\eqref{eq:SWM} to solve the \emph{modified} classical YB equation on the Lie algebra of $G$. The lessons learned from the supercoset case~\cite{Arutyunov:2015qva,Hoare:2015wia,Arutyunov:2015mqj,Borsato:2016ose} suggest that the resulting sigma model will possibly be kappa-symmetric, but that the background fields will probably only solve the equations of generalized supergravity rather than the standard ones.

When applied to classically integrable sigma models, the deformations studied here preserve the integrability. It would be interesting to extend the integrability methods developed in the context of the AdS/CFT correspondence~\cite{Beisert:2010jr,Bombardelli:2016rwb} also beyond the ``abelian'' YB deformations considered so far, namely the ``diagonal abelian'' deformations (considered e.g. in~\cite{Beisert:2005if}  and with an exact spectrum encoded in the equations of~\cite{deLeeuw:2012hp}), and the ``off-diagonal abelian'' deformations (addressed e.g. at one loop in~\cite{Guica:2017mtd}).

\section*{Acknowledgements}
We thank  B. Hoare and S. van Tongeren for interesting and valuable discussions.
RB also thanks the Department of Theoretical Physics and Astrophysics of Masaryk University for hospitality during part of this work.
The work of R.B. was supported by the ERC advanced grant No 341222.

\vspace{2cm}

\appendix

\section{Conventions}\label{sec:conv}
Let us summarize our index conventions in the following table
\be
\begin{aligned}
&\mu,\nu,\ldots: &&\text{labels of all bosonic coordinates}\\
&I,J,\ldots: &&\text{indices of } \alg g \ (\text{the Lie algebra of } G)\text{ and of the dual } \alg g^*\\
&i,j,\ldots: &&\text{labels of coordinates  parameterizing the group } G\\
&M,N,\ldots: &&\text{labels of spectator coordinates, of which }\\
&\qquad m,n,\ldots: &&\qquad \text{labels of bosonic spectator coordinates }\\
&\qquad \u \alpha,\u \beta,\ldots: &&\qquad \text{labels of fermionic spectator coordinates}\\
&A,B,\ldots: &&\text{indices of tangent space, of which }\\
&\qquad a,b,\ldots: &&\qquad \text{indices of bosonic tangent space}\\
&\qquad \alpha,\beta,\ldots: &&\qquad \text{indices of fermionic tangent space}\\
\end{aligned}
\ee

When working with (super)forms we define the components as $A_n=\frac{1}{n!}dz^{M_n}\wedge dz^{M_{n-1}}\ldots \wedge dz^{M_{1}} A_{M_1M_2\cdots M_n}$ and we take the exterior derivative to act from the right, so that $d(A_n\wedge A_m)=A_n\wedge dA_m +(-1)^mdA_n\wedge A_m$.
The (graded) anti-symmetrization of $n$ indices is denoted by $[\cdots]$ and it includes a factor $1/n!$.

\section{An example with local symmetry}\label{sec:example}
To make the discussion in section \ref{sec:local} more concrete we will here apply the rules of NATD to an explicit example with local symmetry (a case also referred to ``with isotropy''). We will follow the discussion in section \ref{sec:local} and show that we reproduce an example worked out in section 4.1 of~\cite{Lozano:2011kb}. The starting point is the $AdS_3\times S^3\times T^4$ background with pure RR flux, and the goal is to apply NATD on the $SO(4)$ global isometry of $S^3$, which has obviously also a local $SO(3)$ symmetry. The metric and the flux are given by
\be
ds^2=ds^2_{AdS_3}+ds^2_{S^3}+ds^2_{T^4},
\qquad
F_3=2\left(vol(AdS_3)+vol(S^3)\right).
\ee
We describe $S^3$ in terms of the coset $SO(4)/SO(3)$, where the generators of $\alg{so}(4)$ satisfy $[J_{ab},J_{cd}]=\delta_{bc}J_{ad}-\delta_{ac}J_{bd}-\delta_{bd}J_{ac}+\delta_{ad}J_{bc}$ and admit the matrix realisation $J_{ab}=E_{ab}-E_{ba}$, in terms of the matrices $(E_{ab})_{cd}=\delta_{ac}\delta_{bd}$. Following~\cite{Lozano:2011kb} we enumerate the generators of the coset part as $T_I=J_{1,I+1}$ where $I=1,2,3$, and the generators of the subalgebra $\alg{so}(3)$ as $T_4=J_{23},T_5=J_{24},T_6=J_{34}$. The metric of the original $S^3$ comes from the piece of the action $\tfrac{T}{2}\int A^I\wedge G_{IJ}* A^J$, where $A=g^{-1}dg,\ g\in SO(4)$ and $G_{IJ}=\text{diag}(1,1,1,0,0,0)$ projects on the coset part of the algebra. We do not need to look at  $AdS_3$ and $T^4$, since the off-diagonal blocks $G_{mI}$ are 0 and therefore the $AdS_3$ and $T^4$ spaces are not affected by the NATD transformations, see~\eqref{eq:tildeGNATD}. It is easy to construct $G_{IJ}-\nu_Kf^K_{IJ}$ that in this case is\footnote{This is the transpose of $M$ of~\cite{Lozano:2011kb}.}
\be
\left(
\begin{array}{cccccc}
 1 & \nu_{4} & \nu_{5} & -\nu_{2} & -\nu_{3} & 0 \\
 -\nu_{4} & 1 & \nu_{6} & \nu_{1} & 0 & -\nu_{3} \\
 -\nu_{5} & -\nu_{6} & 1 & 0 & \nu_{1} & \nu_{2} \\
 \nu_{2} & -\nu_{1} & 0 & 0 & \nu_{6} & -\nu_{5} \\
 \nu_{3} & 0 & -\nu_{1} & -\nu_{6} & 0 & \nu_{4} \\
 0 & \nu_{3} & -\nu_{2} & \nu_{5} & -\nu_{4} & 0 \\
\end{array}
\right),
\ee
and invert it to obtain $N^{IJ}$. Notice that $G_{IJ}$ is not invertible, but we can invert $G_{IJ}-\nu_Kf^K_{IJ}$. For special values of the coordinates $\nu_K$ also $G_{IJ}-\nu_Kf^K_{IJ}$ becomes degenerate. After fixing the gauge, some of these degeneracies will produce singularities in target space.
Taking the symmetric and antisymmetric parts of $N^{IJ}$ we can compute the deformed metric and $B$-field. In the action the contributions are respectively $\tfrac{T}{2}\int d\nu_IN^{(IJ)}*d\nu_J$ and $-\tfrac{T}{2}\int d\nu_IN^{[IJ]}d\nu_J$. These  are still written in terms of all six dual coordinates $\nu_K$, meaning that we should fix the gauge. We fix it as in~\cite{Lozano:2011kb} setting $\nu_1=\nu_2=\nu_6=0$, and we also rename $\nu_3=x_1,\nu_4=x_2,\nu_5=x_3$. In agreement with~\cite{Lozano:2011kb} we find that the $B$-field vanishes and that the metric of the dualised sphere and the dilaton are
\be
\begin{aligned}
ds^2_{\tilde{S}^3}&=\frac{dx_2^2 \left(\left(x_1^2-x_2^2\right)^2+x_2^2
   x_3^2+x_2^2\right)}{x_1^2
   x_3^2}+\frac{\left(x_2^2+x_3^2+1\right)
   dx_3^2}{x_1^2}\\
&+\frac{2 x_2 dx_2 dx_3
   \left(-x_1^2+x_2^2+x_3^2+1\right)}{x_1^2
   x_3}
+\frac{2dx_1}{x_1} \left( x_2 dx_2+ x_3 dx_3\right)+dx_1^2,\\
e^{-2\phi}&=x_1^2x_3^2.
\end{aligned}
\ee
In order to compute the transformation of the RR fields we first need to compute the Lorentz transformation $\Lambda$. Suppose we use labels in tangent space $a=0,\ldots,9$ so that $a=3,4,5$ are the labels for the tangent space of the sphere. Then we can take $E_I{}^a$ to be $E_1{}^3=E_2{}^4=E_3{}^5=1$, and 0 otherwise. Calculating $\Lambda^{ab}=\eta^{ab}-2E_I{}^aN^{IJ}E_J{}^b$ in the above gauge for $\nu_I$ we easily find (for the block with $a,b=3,4,5$) $\Lambda=\text{diag}(1,-1,-1)$. As expected the Lorentz transformation is an element of $SO(1,9)$, since we have dualized an even-dimensional group. In this case it is a simple reflection along $a=4$ and $a=5$. Therefore on spinor indices it is realised just as the product of the two corresponding ten-dimensional gamma matrices. The transformed RR fluxes obtained from $\tilde{\mathcal{S}}=\hat\Lambda \mathcal S$ then agree with the ones of~\cite{Lozano:2011kb}.

\bibliographystyle{nb}
\bibliography{biblio}{}

\end{document}